# Dynamic strain sensing using Doppler-shift-immune phase-sensitive OFDR with ultra-weak reflection array and frequency-tracking

QIANG YANG,[1] WEILIN XIE,[1,2,*] CONGFAN WANG,[1] BOWEN LI,[1] XIN LI,[1] XIANG ZHENG,[1] WEI WEI,[1,2] AND YI DONG[1,2]

[1]*Key Laboratory of Photonics Information Technology, Ministry of Industry and Information Technology, School of Optics and Photonics, Beijing Institute of Technology, No. 5 South Zhongguancun Street, Haidian District, Beijing 10081, China*
[2]*Yangtze Delta Region Academy of Beijing Institute of Technology, Jiaxing 314011, China*
*\*wlxie@bit.edu.cn*

**Abstract:** In distributed fiber-optic sensing based on optical frequency domain reflectometry (OFDR), Doppler frequency shifts due to the changes of disturbances during one sweep period introduce demodulation errors that accumulate along both the distance and time, impairing the sensing performance. Here, we report distributed dynamic strain sensing using Doppler-shift-immune phase-sensitive OFDR based on frequency-tracking and spectrum-zooming with ultra-weak reflection array. Theoretical study has been carried out with the introduction of mismatch coefficient, unveiling quantitatively the impact of Doppler shift. Following a numerical analysis of the proposed method, a retained precision has been experimentally verified regardless of the position mismatch due to the Doppler effect. Doppler-shift-immune sensing for dynamic strains covering continuous spatial resolution over a distance of 1000 m with a 2.5 cm sensing spatial resolution has been demonstrated, verifying the high fidelity promised by the proposed method.

## 1. Introduction

Distributed fiber-optic strain sensing has been widely applied in fields such as traffic facilities [1], aerospace [2], structural health monitoring [3], and biomedicine [4]. When the optical fiber under test (FUT) is subjected to external disturbances, such as strains, the resulting amplitude and phase changes of the Rayleigh backscattering (RBS) along the FUT [5,6] can be utilized by a variety of optical reflectometry techniques for the detection of the disturbances of interest. Regarded as an important tool, optical frequency domain reflectometry (OFDR) [7–11] relying on optical frequency-modulated continuous-wave (OFMCW) based interrogation, has garnered significant interest due to its capability in high spatial resolution and sensitivity at potentially long sensing distances.

Typically, exploiting the response of the RBS in optical fiber [7,12], methods based on either spectral domain [13–15] or phase [9,16] demodulations have been utilized in OFDR. The former is realized by extracting the localized spectral shift between the reference and successive measurement spectra [7,8]. While the latter, also known as phase-sensitive OFDR (φ-OFDR) [17–19], reconstructs the disturbances of interest by detecting the phase changes of the RBS signals. Compared to the sacrificed spatial resolution due to the trade-off for the sensing resolution in the former [20,21], the latter that allows achieving the theoretical spatial resolution defined by the sweep range [9,22], in addition to the high sensitivity earned from the direct phase assessment, has recently drawn a lot of attentions [18,23].

According to the continuous-wave nature of OFDR, throughout an entire sweep period, the OFMCW probe occupies both spatially and temporally along the overall FUT. Such that, strain changes within a sweep period will be captured and then converted into extra phase variations exerted on the RBS from the positions not only in, but also after the strain region all along the fiber [24,25]. As a result, the impact of such time-varying strains within a single sweep period, in the frequency domain, is quite often manifested as Doppler shift. This phenomenon, although similar to that encountered in conventional radar and frequency swept interferometry [26,27], however, imposes much serious impairment as it potentially gives rise to the crosstalk in the context of distributed sensing. The resulting position mismatch due to these unwanted shifts in beat note signals within and after the position where the strain occurs, will inevitably lead to demodulation errors or even failures, thus hampering OFDR measurement.

This way, for strains with a changing rate much smaller than the inverse of the sweep period of the system, they can be regarded as static or quasi-static. Remarkable progresses have been reported when dealing with such slowly-varying or static strain as it can be treated mostly in a stationary fashion [8,23]. However, in scenarios where variations within one sweep period has become non-negligible, such as on-line deformation monitoring for the wings of aircrafts [28] and three-dimensional shape construction in surgical navigation [29], nevertheless, the strain could no longer be regarded as a constant or static but rather as dynamic during a sweep period. In such circumstances where usually a dynamic strain or quantitative vibrations sensing has to be accounted for, Doppler shift has become one dominant limit that must be taken care of.

Intuitively, reducing the sweep period or boosting the sweep rate could alleviate the above issue [17,24]. This, however, not only imposes a limit in sensing distance, but also encounters technical challenges in system complexity such as sweep mechanism and detection bandwidth. Alternative solutions have been studied. One involves estimating the dynamic strain induced frequency shift by the cross-correlation between segments of the RBS trace [25]. The spectral analysis range is then adjusted according to the estimated shift to achieve the spectral tracking for the measurement of multiple vibrations along the fiber. An adaptive two-dimensional cross-correlation for the compensation of the beat frequency shift has been proposed, showing the ability in crosstalk suppression [30]. To date, these methods are almost based on spectral domain demodulation. Thanks to the redundancies gained from the use of the entire spectrum, the underlying extra information could allow for the retrieval of the shift. This is, nevertheless, more challenging in phase demodulation as the RBS within each spatial resolution is treated solely rather than in a collective manner by taking a number of consecutive spatial resolutions as it is in spectral domain demodulation and has gone almost overlooked.

In this paper, Dynamic strain sensing using Doppler-shift-immune φ-OFDR has been proposed and demonstrated. By exploiting the frequency-tracking and spectrum-zooming with ultra-weak fiber grating arrays, it allows a more accurate extraction of the RBS phase at the corresponding beat frequency, thus high precision strain demodulation regardless of the strain variations within single sweep period. The mechanism and impact of Doppler frequency shift induced by the dynamic strain is theoretically studied with a defined metric of mismatch coefficient to characterize quantitatively its impact. Both numerical verification and experiments have been carried out, which effectively demonstrated the Doppler-shift-immune strain sensing with a retained precision even under large mismatch coefficient.

## 2. Concept and operation principle

### 2.1 Doppler frequency shift in φ-OFDR

In a typical φ-OFDR system, the OFMCW signal from the laser source is divided into the probe and reference. By interrogating the FUT with the probe, the beat note signal is obtained through the interference between the yielded RBS and the reference. With $E_0$, $v_0$, and $\gamma$, representing, respectively, the amplitude, initial optical frequency, and frequency sweep rate, of the probe signal, the photo-current of the beat note signal for a certain RBS at position $l_k$ with a round-trip delay of $\tau_k$ within one single sweep period $T_s$ is read

$$I_k(t) \sim 2R(\tau_k)E_0^2 \cos(2\pi\gamma\tau_k t + 2\pi v_0 \tau_k - \pi\gamma\tau_k^2), \quad \tau_k < t \leq T_s, \qquad (1)$$

where $R(\tau_k)$ stands for the reflectivity for the RBS at $l_k$, and $k = 1, 2, 3,...$ refers to the series number for the RBS along the FUT. During the phase demodulation, it is usually considered that $\tau_k$ stays basically stationary within one sweep period. The time-independent Fourier phase at the beat note frequency $f_k = \gamma\tau_k$ is obtained via fast Fourier transform (FFT). It corresponds to the initial phase of the RBS from $l_k$ at the beginning of the sweep period.

When strain $\varepsilon$ is applied on the region with a length $L$ before the position $l_k$ on the FUT, with the strain induced delay change $\tau_\varepsilon$, the beat note signal can be re-written as

$$I_k(t) \sim 2R(\tau_k)E_0^2 \cos\left\{2\pi\gamma(\tau_k+\tau_\varepsilon)t + 2\pi\nu_0(\tau_k+\tau_\varepsilon) - \pi\gamma(\tau_k+\tau_\varepsilon)^2\right\}, \quad \tau_k < t \leq T_s, \qquad (2)$$

In general, the Fourier phase at the resulting beat frequency $\gamma(\tau_k + \tau_\varepsilon)$ can be extracted at the original beat frequency $\gamma\tau_k$, provided $\tau_\varepsilon \ll \tau_k$. With such processing, the strain is attained by differentiating the phases both along the FUT and between successive measurements [18].

The above is valid with the underlying assumption where the strain remains unaltered within each of the sweep periods. For a more general case in which a dynamic strain $\varepsilon(t)$ is considered within one single sweep period, the corresponding time-varying delay change $\tau_\varepsilon(t)$ induced by $\varepsilon(t)$ can be expressed as

$$\tau_\varepsilon(t) = \frac{d\tau_\varepsilon(t)}{dt}\cdot t = \frac{2n\kappa L}{c}\cdot\frac{d\varepsilon(t)}{dt}\cdot t, \quad 0 \leq t \leq T_s, \qquad (3)$$

where $n$, $\kappa$, and $c$, are the refractive index of the fiber, strain coefficient, and the speed of light in vacuum, respectively.

By combining Eqs. (2) and (3) while neglecting the high-order terms since it is rather small compared to the phase consisting of the product between the optical frequency $\nu_0$ and delay, the above equation can be further re-written as

$$\begin{aligned}I_k(t) &\sim 2R(\tau_k)E_0^2 \cos\left\{2\pi\left[\gamma\tau_k + (\gamma t + \nu_0)\cdot\frac{2n\kappa L}{c}\cdot\frac{d\varepsilon(t)}{dt}\right]t + 2\pi\nu_0\tau_k\right\} \\ &= 2R(\tau_k)E_0^2 \cos\left[2\pi(f_k + f_{k,\text{DFS}})t + 2\pi\nu_0\tau_k\right], \quad \tau_k < t \leq T_s,\end{aligned} \qquad (4)$$

where similarly, $\gamma t$ can also be omitted due to fact that $\gamma t \ll \nu_0$ and $f_{\text{DFS}}$ is the corresponding Doppler frequency shift stemming from the time-varying dynamic strain as given by

$$f_{k,\text{DFS}} = \nu_0\cdot\frac{2n\kappa L}{c}\cdot\gamma_\varepsilon, \qquad (5)$$

where $\gamma_\varepsilon = d\varepsilon(t)/dt$ is denoted as the strain slew-rate.

In principle, $f_{\text{DFS}}$ should increase with $\gamma_\varepsilon$ in a linear manner, which is consistent with the Doppler phenomenon in radar system. It can be directly inferred that the presence of Doppler frequency shift will inevitably introduce an unwanted frequency bias in addition to the original beat frequency $f_k$ at $l_k$, leading to a shift to $f_k + f_{k,\text{DFS}}$ for the actually observed beat frequency $f'_k$. According to the relation between the beat frequency and the position in φ-OFDR, where $f_k = \gamma\cdot 2nl_k/c$, such extra frequency shift would probably bring about a mismatch during the positioning along the FUT. Since the phase is still extracted at the original beat frequency $f_k$, it will therefore, severely impair the accuracy for the phase extraction, resulting in probably large errors and even failures during the demodulation.

Obviously, a larger Doppler induced frequency shift, a greater impairment on demodulation. To quantify the degree of offset caused by the Doppler shift with respect to the original beat frequency, here, we define the mismatch coefficient as the ratio between the Doppler induced position shift $z_{\text{DFS}}$ and theoretical spatial resolution $z_{\text{SR}}$, as by definition expressed by

$$\eta_{\text{DFS}} = \frac{z_{\text{DFS}}}{z_{\text{SR}}} = \frac{f_{\text{DFS}}}{\Delta f} = \frac{f_{\text{DFS}}}{f_s}, \qquad (6)$$

where $\Delta f = 1/T_s$ is the original frequency resolution in accordance with the theoretical spatial resolution and equals to the sweep repetition rate $f_s$, i.e. system sampling frequency.

Such relation reveals that for a certain $\gamma_\varepsilon$, a higher $f_s$, i.e. a shorter sweep period $T_s$ it is, a smaller $\eta_{DFS}$, namely a more gentle impact due to the Doppler shift. This can also be intuitively translated into the fact that a fast frequency sweep, i.e. a higher sweep rate $\gamma$, could ultimately alleviate the impact of Doppler effect, which is nevertheless, quite challenging concerning the practical implementation in realizing sufficient fast frequency sweeps.

## 2.2 Issue and method

Concerning the inevitable impairments and limitations, it is therefore, necessary and critical to cope with the Doppler effect without compromising significantly the system. To this end, we firstly depict the influences of the Doppler induced frequency shifts to give a more intuitive physical picture. Exploring the process described in Eq. (4), the RBSs from successive spatial resolutions along the FUT should be obtained at their corresponding beat frequencies.

In the presence of a dynamic strain within a certain region across at least several spatial resolutions, the RBSs carrying the strain information will be shifted in frequency due to the Doppler effect as depicted in Fig. 1(a) for the corresponding intensity traces, frequency shifts after the strain region in proportion to the strain slew-rates, $\gamma_{\varepsilon 1}$ and $\gamma_{\varepsilon 2}$ can be clearly observed. Worth noting that though, rigorously complied with $\gamma_\varepsilon$, the RBSs behavior quite distinctively when within or after the strain region along the FUT. After the strain region, because the strains at different positions within the strain region have already been all accounted for, the RBSs along the FUT shall experience the same delay change given by the integral of the strain within the strain region within one entire sweep period. Therefore, identical Doppler shift should be expected over all the positions after the strain region. To some extent, previous studies that compensates for the Doppler shift through cross-correlation among traces from different sweep periods [25] are primarily relying on this fact.

However, due to the coherent superposition for RBSs with different delay variations within each spatial resolution within the strain region, rather than simply the frequency shift, stochastic fluctuations occur unavoidably on the resulting traces. In consequence, the similarity between traces from different sweep periods has become quite limited. Therefore, it is difficult and even hardly exploited to restore the original beat frequency, and then the actual strain of interest. In addition, the distinct behaviors in RBS responses, respectively, inside and outside the strain region have in turn, led to the strict but annoying requirement [31] that for a quantitative sensing, the two spatial resolutions between which the phase differential is carried out must stay undisturbed, i.e. stable during the sweep period.

Taking all the aforementioned factors into account, the lack of effective access to the actual strain within strain region as a result of the underlying irregular responses, in particular in cases of dynamic strains where the variation of the strain could be hardly neglected within one single sweep period, seriously degrades the precision and further hampers the demodulation of strains. It should be noted that this analysis is not solely limited to φ-OFDR, but also applies to spectral domain demodulation.

To make it possible for extracting the shift and thus restoring the actual strain of interest, in the proposed method, we exploit fiber embedded with ultra-weak reflection grating arrays as the FUT. Benefiting from the relatively dominant reflectivity thus the higher signal-to-noise ratio (SNR) compared with the regular RBS, this in principal allows equivalently imprinting onto the sensing fiber a signature so as to provide an opportunity for the access to the possible frequency shifts. Without scarifying the spatial resolution, the spacing of the embedded grating is set equal to the theoretical spatial resolution $z_{SR}$ defined by the sweep range. Moreover, to cope with the limited accuracy in phase extraction due to the limited frequency resolution, by incorporating a digital zero-padding enabled spectrum zooming, it should enable a high fidelity strain reconstruction with the high precision frequency tracking for and the phase extraction from the shifted beat frequencies.

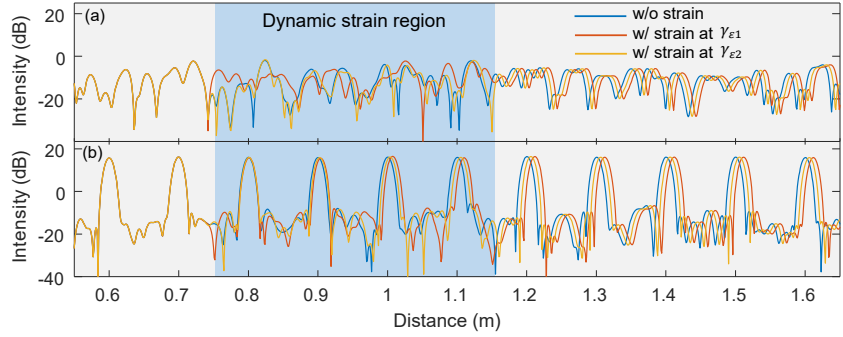

Fig. 1. The intensity traces of (a) RBS and (b) weak reflection array in the cases of without and with strain.

As conceptually sketched in Fig. 1(b), the Doppler frequency shift in weak reflection array increases continuously in an accumulative manner along the distances within the strain region, while it remains constant when beyond the strain region. With the weak reflection induced fiber signature and the enhanced frequency resolution, it is obvious that such accumulated shifts for the beat frequencies associated to the successive spatial resolutions can be steadily tracked at the corresponding reflection points not only inside but also outside the strain region.

### 2.3 Numerical verification

Numerical study is carried out to explore quantitatively the impact of the Doppler shift and the effectiveness of the proposed system. The frequency sweep range and period are 10 GHz and 1 ms, corresponding to a 1 kHz and 1 cm theoretical frequency resolution and spatial resolution, respectively. Taking into account the broadening of the beat note spectrum due to the applied Hanning window for the suppression of the sidelobes during FFT, a 2.00 m grating embedded fiber with a 3 cm spacing between adjacent weak reflection points acts as the FUT. Considering more universally when a dynamic strain in a sinusoidal manner is taken as described by $\varepsilon(t) = \varepsilon_0 \sin(2\pi f_\varepsilon t)$, where $\varepsilon_0$ and $f_\varepsilon$ are the amplitude and frequency of the strain, respectively. Then, taking with Eq. (5), Eq. (6) can be re-written as

$$\eta_{\text{DFS}} = \frac{4\pi n v_0 \kappa}{c} \cdot L\varepsilon_0 \cdot \frac{f_\varepsilon}{f_s} \cdot \cos(2\pi f_\varepsilon t). \qquad (7)$$

According to the time-varying $\gamma_\varepsilon(t)$ over a complete sinusoidal period, the resulting Doppler shift, thus $\eta_{\text{DFS}}$, should change consequently and maximizes when at the zero-crossing point with a maximum $\eta_{\text{DFS}} = 4\pi n v_0 \kappa L\varepsilon_0 f_\varepsilon / (f_s c)$. In the following, a sinusoidal strain at 10 Hz with an amplitude of 20 με is applied within the region from 0.50 m to 1.50 m.

Phase extraction at weak reflections is analyzed in case for consecutive spatial resolutions. Compared with the traces obtained with and without strain as shown in Fig. 2(a) when at the maximum $\eta_{\text{DFS}}$, a manifested shift from 120 kHz ($f_{k,A}$) to 121 kHz ($f_{k,B}$) can be observed for the beat frequency of $k^{\text{th}}$ weak reflection located at 1.20 m. Obviously, it becomes impossible to extract the phase at the original beat frequency $f_{k,A}$, as it has been drastically inconsistent with what it is supposed to be in the absence of strain due to the Doppler effect. Meanwhile, as the analysis window in FFT is theoretically dictated by the sweep period in OFDR, the amplitude and phase can only be acquired at the integer multiples of theoretical frequency resolution (1 kHz in this analysis). On the basis of FFT, rather than staying unaltered, the Fourier phase varies in a $2\pi$ range along the envelope of the beat note spectrum when within the broadened frequency resolution accounting for the applied window. Such that, it inevitably leads to significant errors when the actual beat frequency deviates and hardly coincides at those dedicated frequencies [32]. In consequence, even by tracking the shifted beat frequency, phase and amplitude can be only accessed at those dedicated frequencies separated by the theoretical frequency resolution, with however, a mostly degraded precision in this case.

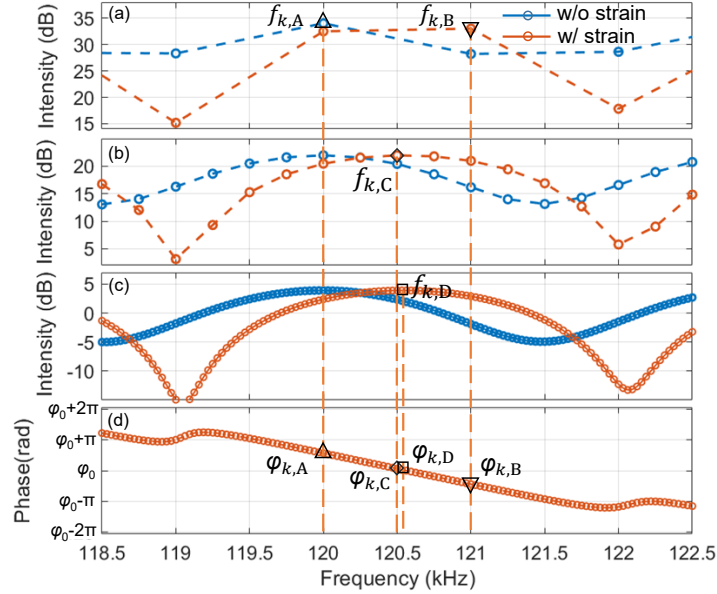

Fig. 2. The intensity traces of weak reflection during the interrogations of without strain (blue) and with dynamic strain (red), in the cases of (a) without spectrum-zooming, (b) with 4-times zero-padding, and (c) with 32-times zero-padding. (d) The phase spectrum under the condition of (c) with dynamic strain.

To track more accurately the shifted beat frequency and access more precisely the amplitude and phase, the results in case of spectrum-zooming by employing 4 and 32 times zero-padding are given in Fig. 2(b) and 2(c), respectively. It clearly shows that a more accurate positioning and thus tracking for the shifted beat frequency is permitted thanks to the improved frequency resolution. With a higher padding density, the tracking with a higher precision is expected as the demodulation at 120.5625 kHz ($f_{k,D}$) in the latter case follows more closely the designated Doppler shift of 0.5625 kHz and is in accordance with the given $\eta_{DFS} = 0.28$. The associated phase spectrum in Fig. 2(d) verifies the promised more accurate extraction for Fourier phases $\varphi_{k,D}$ at $f_{k,D}$ with the phase error being only ~0.0066 rad. It confirms a significant improvement compared with the non-trivial phase error of -1.3878 rad without spectrum zooming, not to mention that of ~1.7845 rad in the absence of frequency-tracking. It is noted that aiming at a high precision, a denser zero-padding is always preferred however, at the expense of increased processing consumption. An adequate spectrum zooming should take into account the trade-off between the complexity and the precision limit governed by the system noise floor.

The variations of the phase error with respect to the mismatch coefficient $\eta_{DFS}$ are further assessed under different conditions as shown in Fig. 3(a). For the demodulation in conventional φ-OFDR, i.e. without frequency-tracking (blue curve), along the obvious degradation in phase error with the increase of $\eta_{DFS}$, one can also observe an irregular but periodic variation in phase error. Such trend is mainly attributed to the nonlinear phase evolution along with the increase of $\eta_{DFS}$. Significant reduction in the phase error can be achieved by implementing frequency-tracking (yellow, purple and red curves). Nevertheless, the periodic variations in the resulting phase error has been retained in all cases because the deviation, thus accuracy for the tracking of the beat frequency also varies regularly with respect to $\eta_{DFS}$. The higher frequency resolution gained from the increased density for the zero-paddings improves the phase error performance as manifested by the significantly reduced standard deviation as low as 0.0498 rad. This verifies the effective mitigation for the adverse effects of the Doppler shift on phase demodulation.

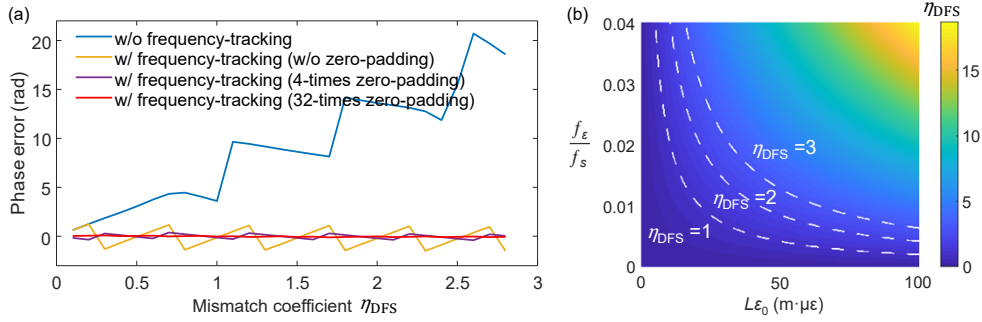

Fig. 3. (a) The variations of phase error with mismatch coefficient $\eta_{DFS}$ under different conditions. (b) The maximum $\eta_{DFS}$ as a function of $L\varepsilon_0$ and $f_\varepsilon/f_s$ under applied sinusoidal strain.

According to Eq. (7), $\eta_{DFS}$ is determined by the contribution both from the strain amplitude and frequency. These two factors are weighted by the metrics $L\varepsilon_0$ and $f_\varepsilon/f_s$ concerning the strain induced delay changes and the ratio between the strain frequency to the inverse of the sweep period (known as the system sampling frequency $f_s$ in OFDR), respectively. This way, the maximum $\eta_{DFS}$ for strains of certain amplitude and frequency, which is found at the zero-crossings for the sinusoidal strain, as a function of these two factors is outlined in Fig. 3(b). The indications when $\eta_{DFS} = 1$ describes the cases where the position shift corresponding to the Doppler shift exactly reaches one theoretical spatial resolution. This has been preliminarily considered as the limit [25,33] while a recent study has shown that significant crosstalk occurs when it is above 0.32 [30] for spectral domain demodulation.

It is evident that $\eta_{DFS}$ can easily exceed 1 for typical configurations from a practical point of view. In general, the measurable strain frequency must retain within half of $f_s$ according to the Nyquist sampling theorem. A more rigorous criterial for the measurable strain frequency should however, take into account the effect of Doppler shift. Since $\eta_{DFS}$ increases with the frequency of the strain for a certain fixed $f_s$, a lower bound that is often far below the theoretical Nyquist limit of $f_s/2$ should be taken, unless $L\varepsilon_0$ is particularly small. In addition, $\eta_{DFS}$ increases with the length of strain region as well as the strain amplitude. This indicates that when a continuous strain is applied over more than one spatial resolution, the Doppler shift will be accumulated along the strain region, as well as $\eta_{DFS}$. It eventually leads to a more severe demodulation error and even failures along the strain region. Therefore, the adaptation of the proposed scheme is of particular importance to immunize against the influence of Doppler shift.

## 3. Experiment and results

### 3.1 Experimental setup

The experimental setup is shown in Fig. 4. The output from a fiber laser is externally modulated by a phase modulator (PM) driven by a linearly frequency swept signal generated from an arbitrary waveform generator (AWG). After selecting the desired sideband through an optical band-pass filter (OBPF), the obtained OFMCW probe with a sweep range of 12.5 GHz and a sweep period of 1 ms is sent to the OFDR interferometer for sensing. The FUT consists of a 1 km long single-mode fiber (SMF) and an ultra-weak fiber-grating array with a 2.5 cm grating spacing and ~-40 dB reflectivity. The 60-m portion at the tail of the SMF is wound around a piezoelectric transducer (PZT) to apply dynamic strains. In addition, a small PZT stack (PZTS) with a length of 2.0 cm, consisting of multiple piezoelectric chips that are stacked face to face, is adhered to the grating fiber to apply axial strain [31]. After that, a motor- driven fiber stretcher is installed to apply continuously distributed strains. The beat signal from the interferometer is received by a balanced photodetector (BPD) and sent to an oscilloscope for post processing.

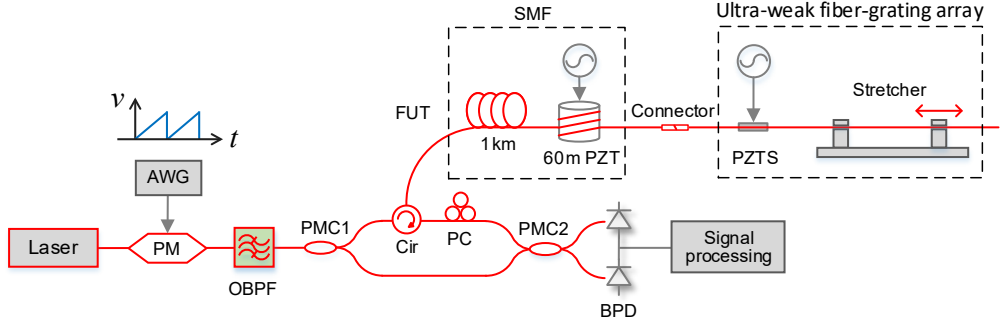

Fig. 4. Experimental setup. AWG, arbitrary waveform generator; PM, phase modulator; OBPF, optical band-pass filter; PMC, polarization-maintaining coupler; Cir, optical circulator; FUT, fiber under test; SMF, single-mode fiber; PZT, piezoelectric transducer; PZTS, piezoelectric transducer stack; PC, polarization controller; BPD, balanced photodetector.

*3.2 Results, analysis, and discussion*

In accordance with the verification in the numerical study, we firstly evaluate the ability of the proposed method in case of strain within one single spatial resolution. The strain of interest is introduced by the PZTS located at 1070.02 m, before which the fiber-wound PZT is positioned at 1008.72 m to emulate the occurrence of Doppler shift by applying a fast-varying dynamic strain. A sinusoidal excitation with an amplitude of 100 $V_{pp}$ and a frequency at 20 Hz is applied to the PZTS while a sinusoidal strain of 5 $V_{pp}$ at 8 Hz is exerted on the PZT at the meantime. Accounting for the length and response of the PZT, it would in principle lead to a Doppler shift of a maximum $\eta_{DFS}$ of 3.1 for all the positions along the FUT after this PZT. Above all, the demodulated strain when without the excitation on the PZT, i.e. without Doppler shift, is given in green dashed line in Fig. 5(a) for reference. In contrast, distortions in terms of abrupt spurs and burrs with a drift along time can be observed due to the position mismatch stemming from the obvious Doppler shifts within the strain region induced by the PZT. The dramatic errors and its time-varying property, as manifested as the drift in almost a non-stationary behavior, will probably result in failures in traditional demodulation.

By employing frequency-tracking in connection with spectrum-zooming with 8-times zero-padding, the result (red solid line) is basically consistent with that when without Doppler shift, indicating the effective recovery of the actual strain of interest. In the associated power spectral densities (PSDs), a significantly improved sensitivity, with an increase of 30.6 dB for the SNR at 20 Hz compared with other cases has also been verified. Worth emphasizing that identical harmonic behaviors, such as the sidebands at 40, 60 and 80 Hz, are observed in both cases when with and without Doppler shift. It means that the nonlinearity is essentially inherited from the original irregular response of the PZTS and the system response rather than additionally from the proposed method, confirming its high fidelity.

The precisions with different demodulation schemes in the presence of the Doppler shift are further compared. By selecting a non-strain fiber segment of 1.00 m right after the strain region consisted of the PZT, the standard deviation of the demodulated strains that is also equivalently regarded as the averaged sensitivity, obtained in 500 interrogations is taken as the metric for the precision. The results under $\eta_{DFS}$ of 3.1 when without and with the frequency-tracking are shown, respectively, in Figs. 6(a) and 6(b). For the former, inevitable errors and even failures probably take place due to the Doppler induced positioning mismatch. As the phase differential and unwrapping is to be carried out relying on the incorrect phases obtained at these inaccurate beat frequencies, the resulting strain errors will be accumulated both temporally and spatially, leading to a deteriorated precision at positions even after the strain regions. For instance, a more serious degradation along the measurement time can be identified in Fig. 5(a). It behaviors in a similar manner for the results attained from other regions after the PZT induced strain region, constituting a cross-verification for the impairing error accumulation. In contrast, as the phases can be extracted at a more accurate beat frequency with the frequency-tracking similar to that in the strain region, a significantly reduced variation of the strain over the non-strain region can also be steadily found as inferred by Fig. 6(b).

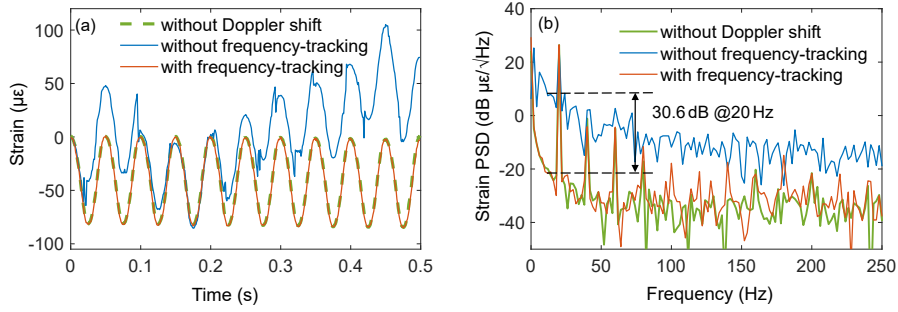

Fig. 5. (a) Demodulated strains without Doppler shift (green dashed line), without frequency-tracking under Doppler shift (blue solid line), and with frequency-tracking under Doppler shift (red solid line), respectively, at the PZTS located at 1070.02 m. (b) Corresponding power spectral densities.

Furthermore, the strain precision at a varied $\eta_{DFS}$, i.e. a varied Doppler shift, by applying different amplitudes to the PZT, are quantitatively evaluated. As shown in Fig. 6(c), when $\eta_{DFS}$ stays roughly smaller than ~0.7, the precision degrades in a very slow pace regardless of the demodulation schemes, even without frequency-tracking. However, as it becomes greater, a steep drop along with a random variation can be found when without frequency-tracking. In fact, an increase both in strain amplitude and frequency will lead to a rise in the peak value of $\eta_{DFS}$, while strain frequency also determines the instantaneous change of $\eta_{DFS}$. Interestingly, in cases of frequency-tracking with different scales of spectrum-zooming, merely a slight discrepancy in the deteriorations can be discovered with the increase of $\eta_{DFS}$. Compared to the case when without frequency-tracking, the strain precision under tracking with 32-times zero-padding has been improved by ~13 dB, as shown by the purple curve in Fig. 6(c). The corresponding comparison as a function of scales of spectrum-zooming is summarized in Fig. 6(d), indicating that a 4-times zero-padding is quite sufficient to achieve a significant improvement in strain precision. Since large scales of zero-padding may compromise the computation consumption, this relation indicates the boundary for an efficient performance enhancement can be achieved with acceptable trade-off, which is important from a practical point of view. In addition, in some application scenarios where only the local position of the fiber needs to be focused on, other methods such as Zoom-FFT algorithm [34] and Chirp-Z algorithm [35] can be employed without significantly increasing the processing consumption.

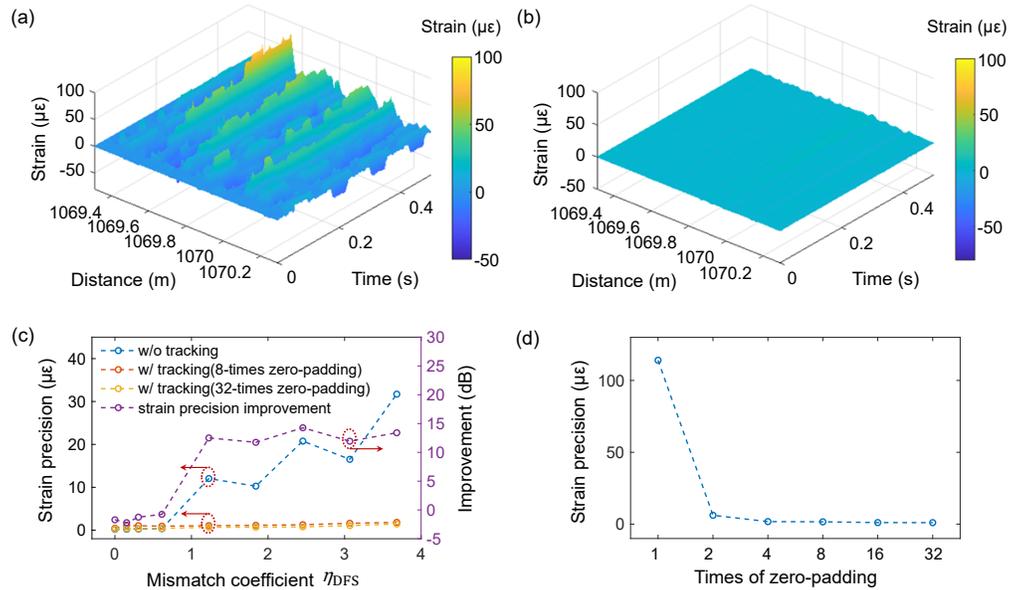

Fig. 6. Results obtained in the non-strain fiber segment after the strain region when (a) without and (b) with the proposed frequency-tracking, respectively. Demodulated strain precisions at (c) different $\eta_{DFS}$ and (d) different times of zero-padding with a fixed $\eta_{DFS}$ of 3.1, at a distance of ~1000 m.

The capability for distributed sensing in a fully continuous manner is experimentally verified by applying dynamic strains via stretching the motor-driven fiber stretcher. A 0.5 m segment of grating fiber is clamped by the fixtures on the moving stages of the stretcher. The gauge length is set to 2.5 cm, which is consistent with the spacing of the gratings. A sinusoidal stretch is applied on the moving stages with a calibrated magnitude 175.8 με at 5 Hz. The demodulated strain without frequency-tracking shows severe distortion due to position mismatch induced phase demodulation error and failure as testified in Fig. 7(a). Similar to the results in Fig. 6(a), the fiber in the non-strain region after the strain region is also contaminated by the accumulated phase error. On the contrary, by using frequency-tracking with 8-times spectrum-zooming, the actual strain can be steadily recovered as given in Fig. 7(b), where the precision remains almost consistent before, inside, and even after the strain region.

For further analysis, the applied strain and $\eta_{DFS}$ with different measurement times at the tail of the strain region are shown in Fig. 7(c). $\eta_{DFS}$ evolves in proportion to the instantaneous slew-rate $\gamma_\varepsilon(t)$ of the applied strain, that is, its derivative. As $\eta_{DFS}$ changes periodically over time, the measurements at the minimum ($\eta_{DFS} = 0$) and the maximum Doppler shifts ($|\eta_{DFS}| = 1.2$) are marked. Different from the periodically behavior of $\eta_{DFS}$, the precision within the strain region deteriorates continuously over time when without frequency-tracking as shown in Fig. 7(d) due to accumulation in phase errors. Meanwhile, the demodulated strains also experience distortion because of the phase errors from the preceding measurements, where the impact of Doppler shift are accumulated temporally. In contrast, the strain precision remains almost constant when frequency-tracking is implemented.

For further insights, the details at the marked values of $\eta_{DFS}$ in Fig. 7(c) are analyzed. The strain at 5th (5 ms), 105th (105 ms), 205th (205 ms), and 305th (305 ms) measurement in Fig. 7(a) are sketched in Fig. 7(e), respectively. Though with a minimum $\eta_{DFS} = 0$ so that the Doppler shift is supposed to be negligible, nevertheless, strain demodulation is inevitably contaminated in the latter three measurements due to the accumulation of phase errors over time. It is not surprising that a gradual deterioration in a similar manner should be found when the absolute value of $\eta_{DFS}$ is maximized as it reaches a peak of 1.2 as seen in Fig. 7(f). Conversely, by exploiting the frequency-tracking, a high fidelity demodulation can be accomplished regardless of $\eta_{DFS}$. As shown in Fig. 7(g), the demodulated strains at the 5th (blue curve) and 205th (yellow curve) measurements are close to zero, consistent with the applied strain as shown in Fig. 7(c). The maximum magnitude of the demodulated strain (at 105th or 305th measurements) is ~180.5 με with merely 2.67% relative error with respect to the applied value, indicating a high reliability of the propose method. In addition, if the distance corresponding to the rising edge of 10-90% is taken as a more rigorous sensing spatial resolution, it is approximately 3.93 cm in this sensing system. It should be noted that in case of arbitrary strains, no matter single or multiple, occurring before a certain grating, other strains occur after this grating can be readily obtained owing to the differential between the adjacent grating during the demodulation process.

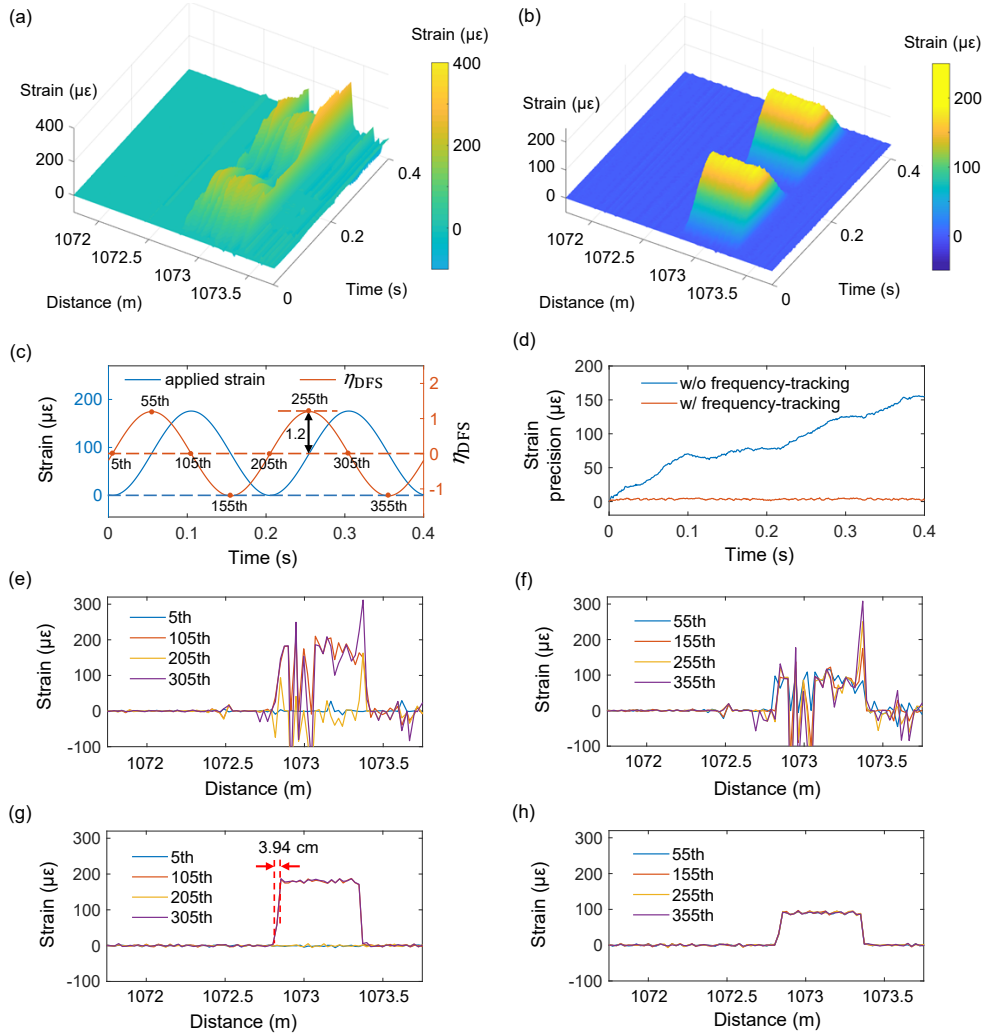

Fig. 7. Demodulated continuously distributed strains in the cases of (a) without frequency-tracking and (b) with frequency-tracking. (c) The applied strain and mismatch coefficients $\eta_{DFS}$ at different measurement times at the tail of the strain region. (d) The variations of strain precision within the strain region with measurement times. (e) (f) Demodulated strains at $\eta_{DFS} = 0$ and $|\eta_{DFS}| = 1.2$ in (a), respectively. (g) (h) Demodulated strains at $\eta_{DFS} = 0$ and $|\eta_{DFS}| = 1.2$ in (b), respectively.

Worth noting that some common issues in distributed fiber-optic sensing systems that rely on these kinds of discrete reflection points, here fiber grating for instance, should be taken care of in field implementations. It is undeniable that there is always a trade-off between the possible number of reflection points and the reflectivity. Since a larger reflectivity will result in a more significant return loss, therefore, though with an ultra-weak reflectivity, this still puts forward a stringent limit on the achievable sensing distance. Optimal solution should consider both the number and reflectivity according to the actual application requirements. Besides, the inevitable additional deployment brings about difficulties for the sensing through now existing fiber links or communication fiber system using standard SMF. Though for many emerging scenarios, the deployment can be accomplished during the fabrication or assemble [28, 29], in cases when standard SMF cannot be easily replaced, more general and flexible techniques are still required.

## 4. Conclusion

In this paper, we have investigated the Doppler frequency shift in φ-OFDR system, and proposed the method of frequency-tracking to immune its adverse effect. The strain change within a single sweep period leads to Doppler shift, causing position mismatch in different interrogations. This introduces demodulation errors that accumulate along both the distance and time, thus impairing the sensing performance. In the context of mismatch coefficient which is defined to characterize the relative effect of Doppler shift, the phase errors of frequency-tracking and spectrum-zooming are evaluated by theoretical simulation, indicating accurate phase extraction by tracking the shifted beat frequency. More importantly, Doppler-shift-immune strain sensing with retained precision is experimentally demonstrated regardless of the position mismatch, and the capability for the high-fidelity demodulation is verified in distributed sensing of continuous strains covering a distance of 1000 m with a 2.5 cm spatial resolution. The proposed method holds wide application prospects for multi-event or continuous distributed dynamic strain sensing with high spatial resolution and high sensitivity.

**Funding.** National Natural Science Foundation of China (62322503, 61827807).

**Disclosures.** The authors declare no conflicts of interest.

**Data availability.** The data underlying the results presented in this paper are not publicly available at this time but may be obtained from the authors upon reasonable request.